\documentclass[11pt]{article}

\usepackage[final]{acl}

\usepackage{times}
\usepackage{latexsym}

\usepackage[T1]{fontenc}

\usepackage[utf8]{inputenc}

\usepackage{microtype}

\usepackage{inconsolata}

\usepackage{graphicx}

\usepackage{booktabs}
\usepackage{amsmath}
\usepackage{tabularx}
\usepackage{listings}
\usepackage{xcolor}
\usepackage{multirow}
\usepackage{placeins}

\title{Jamendo-MT-QA: A Benchmark for Multi-Track Comparative Music Question Answering}

\author{
\textbf{Junyoung Koh}\textsuperscript{1,2,6}\thanks{Corresponding author: \texttt{solbon1212@yonsei.ac.kr}} \quad
\textbf{Jaeyun Lee}\textsuperscript{2,3} \quad
\textbf{Soo Yong Kim}\textsuperscript{6} \quad
\textbf{Gyu Hyeong Choi}\textsuperscript{5,6} \\
\textbf{Jung In Koh}\textsuperscript{} \quad
\textbf{Jordan Phillips}\textsuperscript{4,6} \quad
\textbf{Yeonjin Lee}\textsuperscript{1} \quad
\textbf{Min Song}\textsuperscript{7} \\
\\
\textsuperscript{1}Yonsei University \quad
\textsuperscript{2}KRAFTON \quad
\textsuperscript{3}University of Oxford \\
\textsuperscript{4}George Mason University \quad
\textsuperscript{5}Sungkyul University \quad
\textsuperscript{6}MODULABS MAAP \quad
\textsuperscript{7}Onoma AI 
}

\begin{document}
\maketitle
\begin{abstract}
Recent work on music question answering (Music-QA) has primarily focused on
single-track understanding, where models answer questions about an individual
audio clip using its tags, captions, or metadata.
However, listeners often describe music in comparative terms, and existing
benchmarks do not systematically evaluate reasoning across multiple tracks.
Building on the Jamendo-QA dataset, we introduce Jamendo-MT-QA, a dataset and
benchmark for \emph{multi-track comparative question answering}.
From Creative Commons-licensed tracks on Jamendo, we construct \textbf{36,519}
comparative QA items over \textbf{12,173} track pairs, with each pair yielding
three question types: yes/no, short-answer, and sentence-level questions.
We describe an LLM-assisted pipeline for generating and filtering comparative
questions, and benchmark representative audio--language models using both
automatic metrics and LLM-as-a-Judge evaluation.
\end{abstract}

\section{Introduction}

Music is often experienced relationally: listeners compare tracks by mood, energy, instrumentation, or stylistic similarity (e.g., “this song feels darker than the previous one”).
Despite this, most existing music benchmarks \cite{musiceval,Marble}
focus on single-item tagging \cite{MagnaTagATune}, captioning \cite{doh2023toward,AudioCaps,MusicLM}, or classification \cite{GTZAN,FMA,fma_challenge}, and do not directly test a model's ability to reason \emph{across} multiple tracks.
Recent analyses further suggest that performance on some Music-QA benchmarks may be driven by text-based cues rather than genuine audio perception,
highlighting the need for evaluation settings that explicitly probe reasoning over audio inputs \cite{areyoureallylistening}.

In parallel, large language models (LLMs) and audio-language models \cite{AudioPaLM,AudioCaption} have achieved strong results on textual QA and multimodal understanding.
Yet it remains unclear how well these systems support \emph{comparative} music reasoning, where a model must integrate information from two or more tracks and express the relation in natural language.

Jamendo-QA \cite{jamendoqa} recently introduced a large-scale dataset of QA pairs and captions aligned with Jamendo music audio, targeting general music understanding.
Building on this resource, we propose Jamendo-MT-QA, a dataset for \textbf{\underline{M}ulti-\underline{T}rack comparative \underline{Q}uestion \underline{A}nswering}.
Each data point corresponds to a \textit{track pair}, for which we generate three comparative questions (yes/no, short-answer, and sentence-level), collectively forming a \textit{QA group}.

Beyond dataset construction, we empirically investigate the difficulty of
multi-track comparative question answering by benchmarking several recent
audio--language models.
Our evaluation reveals that, despite strong performance on single-track
audio understanding, existing models struggle to generate well-grounded
sentence-level comparisons across multiple tracks.
These findings motivate the need for a dedicated benchmark that explicitly
targets comparative music reasoning.

\paragraph{Contributions.}
Our main contributions are:
\begin{itemize}
    \item We construct Jamendo-MT-QA on top of Jamendo-QA, comprising
\textbf{36,519} comparative QA items over \textbf{12,173} track pairs,
covering genre, mood, instrumentation, and production-related attributes.
    \item We design an \textbf{LLM-assisted pipeline} for generating and filtering three comparative questions per track pair (yes/no, short-answer, sentence-level).
    \item We provide a \textbf{baseline benchmark} of representative multi-audio and caption-based audio-language models under a unified evaluation protocol, using both automatic metrics and LLM-as-a-Judge evaluation.
\end{itemize}

\section{Related Work}

\begin{figure*}[t]
  \centering
  \includegraphics[width=\textwidth]{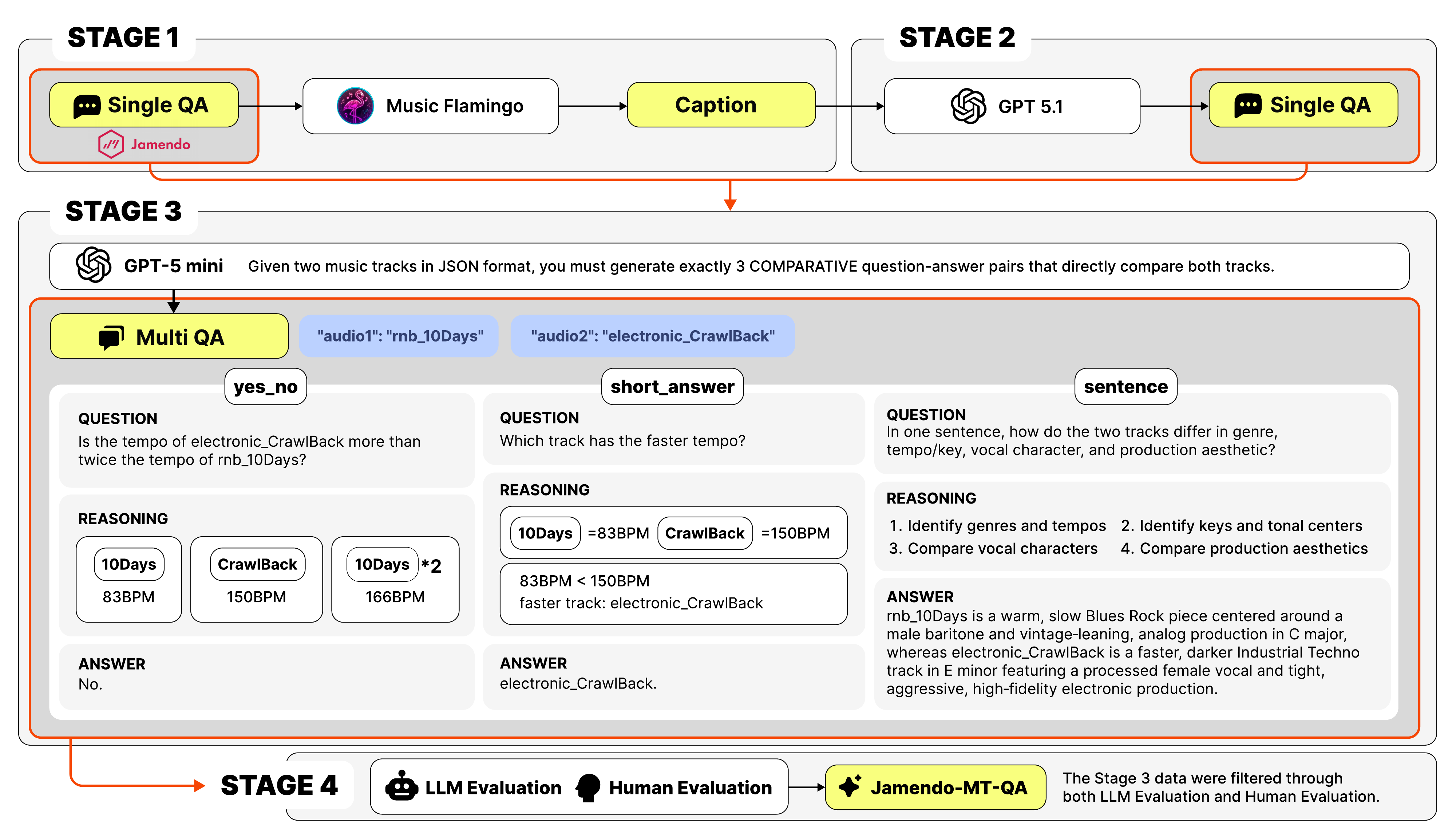}
  \caption{
Overview of the Jamendo-MT-QA construction pipeline.
Starting from the Jamendo-QA dataset, we first generate
rich single-track captions using Music Flamingo, expand them into single-track QA pairs using GPT-5.1, and construct comparative QA groups consisting of three question types (yes/no, short-answer, sentence-level) for each track pair via an LLM-assisted generation pipeline.
Finally, we apply human evaluation and LLM-as-a-Judge assessment for quality control and filtering, resulting in the final Jamendo-MT-QA dataset.}
  \label{fig:jmt_qa_pipeline}
\end{figure*}

\paragraph{Music-QA and Captioning.}
Prior datasets such as MUSIC-AVQA \cite{musicavqa}, MusicQA \cite{mqad}, and MusicXQA \cite{musicxqa} focus on answering questions about music or music-related video, often using symbolic representations or constrained QA formats.
Other large-scale resources, including MTG-Jamendo \cite{mtgjamendo}, JamendoMaxCaps \cite{jamendomaxcaps}, and LP-MusicCaps \cite{lpmusiccaps}, provide tags or captions but not explicit QA pairs.
Jamendo-QA \cite{jamendoqa} dataset addresses Music-QA at scale for single tracks; our work extends this direction to multi-track comparative reasoning.

\paragraph{Multimodal Audio-Language Models and Benchmarks.}
Audio-language models based on contrastive learning, such as CLAP \cite{elizalde2023clap}
and MuLan \cite{mulan}, align music or audio with natural language in a shared
representation space and have shown strong performance on retrieval and captioning \cite{AIBA}.
Beyond representation learning, encoder--decoder architectures and
audio-conditioned LLMs \cite{qwen2audio,GAMA,SALMONN} enable more flexible
audio--text generation and understanding.
Recent models such as MU-LLaMA \cite{mullama,mumullama},
Music Flamingo \cite{musicflamingo}, and ChatMusician
\cite{chatmusician} further extend this line of work by supporting
instruction-following and interactive reasoning grounded in music audio.

In parallel, several benchmarks have been proposed to evaluate audio-language
models across diverse tasks and domains.
The MAE benchmark \cite{beyondsingleaudio} evaluates multi-audio processing
capabilities across speech and general sound domains.
AIR-Bench \cite{airbench} introduces an instruction-following benchmark for
audio-language models spanning speech, environmental sounds, and music.
AudioBench \cite{audiobench} further provides a broad, task-diverse evaluation
suite for AudioLLMs, covering captioning, question answering, retrieval, and
reasoning over general audio. These benchmarks primarily assess general audio understanding or the ability
to process multiple audio inputs, typically in single-track or non-comparative
settings.

In contrast, Jamendo-MT-QA specifically targets \emph{comparative reasoning}
between music tracks, requiring models to integrate perceptual attributes across
tracks and generate structured comparative answers.

\paragraph{Comparative and Relational Question Answering.}
Comparative question answering has been extensively studied in the NLP domain,
particularly in multi-hop and relational QA settings \cite{MoreHopQA, MEQA,lee2026automaticinterdocumentmultihopscientific}.
Representative benchmarks such as HotpotQA \cite{hotpotqa} and WikiMultiHopQA
\cite{wikimultihop} require models to aggregate evidence across multiple documents,
while DROP \cite{DROP} emphasizes discrete and logical reasoning over textual
contexts.

Recent work on audio-language benchmarks has highlighted that high accuracy on existing QA tasks can mask deficiencies in genuine multimodal perception and reasoning.
For example, in Music-QA, text-only models can achieve strong results even without access to audio inputs, suggesting dataset biases and shortcut cues rather than true perceptual inference \cite{areyoureallylistening}.
Such diagnostic perspectives motivate benchmarks that more explicitly probe relational and reasoning capabilities beyond surface-level pattern matching.

However, existing analyses in the music and audio domain have primarily focused on single-track understanding or perceptual grounding in isolation \cite{musicbert,CLAMP3}.
They have not systematically examined \emph{comparative} or \emph{relational} reasoning across multiple audio inputs, where models must integrate perceptual attributes, temporal structure, and cross-track relationships.
Our work draws inspiration from both multi-hop QA and recent diagnostic benchmark analyses by formulating comparative Music-QA as a multi-track reasoning problem, and by providing a dataset and benchmark that explicitly require integrating information across multiple music tracks.

\begin{figure*}[t]
\centering
\includegraphics[width=0.95\textwidth]{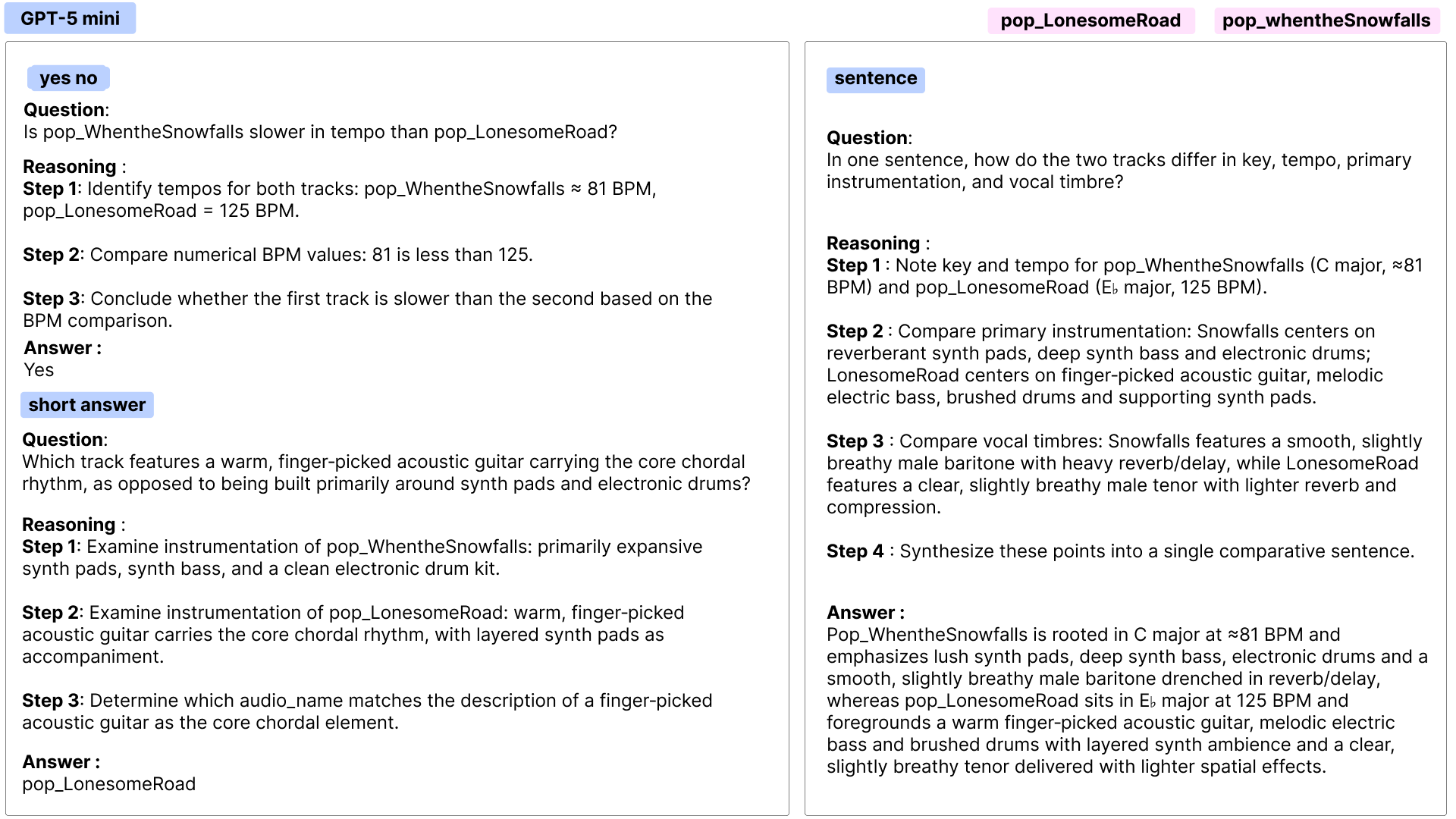}
\caption{
Qualitative example of a Stage~3 comparative QA group generated by \textbf{GPT-5 mini}
for the same track pair. The generator produces three comparative question types
(yes/no, short-answer, sentence-level) and provides explicitly step-by-step
reasoning, which improves interpretability and makes the \textit{reasoning} field
more suitable as a supervision signal for downstream training.
}
\label{fig:stage3_gpt5mini}
\end{figure*}

\section{Jamendo-MT QA Dataset}
\label{sec:dataset}

\paragraph{Base Dataset: Jamendo-QA.}
Jamendo-QA \cite{jamendoqa} is built from Creative Commons-licensed tracks on the Jamendo platform and provides question–answer pairs and captions aligned with music audio. Each sample links an audio file, rich metadata (genre, tempo, gender, etc.), and automatically generated QA pairs. The dataset spans a wide range of genres and musical attributes. In this work, we treat Jamendo-QA as the foundation for our multi-track comparative QA dataset.

\subsection{Stage 1: High-quality music captioning}
For each track,   structured natural language description using Music Flamingo \cite{musicflamingo}.
Given a raw audio, Music Flamingo produces detailed captions covering genre, tempo, key, instrumentation, production style, vocal characteristics, and lyrical themes. These captions form a dense semantic representation of each track that serves as the basis for downstream QA generation.
To ensure audio--text consistency, we further conduct a human evaluation in which annotators with access to the underlying audio assess the alignment between the generated captions and the musical content. The detailed evaluation method and results are provided in Appendix~\ref{sec:human-evaluation}

\subsection{Stage 2: Single-track QA expansion}
We then expand each caption into multiple high-quality single-track QA pairs using GPT-5.1. The model generates diverse factual and descriptive questions about the track (e.g., genre, tempo, key, instrumentation, mood), producing an enriched single-track QA resource.

\subsection{Stage 3: Multi-track Comparative QA Generation}
To construct comparative QA pairs, we combine two track descriptions (caption, metadata, and single-track QA) into a structured JSON input and prompt the model to act as a \emph{music comparison expert}. In this stage, we use \textbf{GPT-5 mini} as the primary generator for dataset construction.
For each track pair, the model produces exactly three comparative questions corresponding to yes/no, short-answer, and sentence-level formats.
All questions are required to explicitly reference both tracks.
The full output schema is provided in Appendix~\ref{sec:output-schema}.

We experimented with several alternative LLMs for Stage~3
generation, including Qwen3-32B \cite{qwen3}, InternLM3-8B \cite{internlm2}, Gemma 3 12B \cite{Gemma3}, and Claude 4.5 Haiku \cite{claude45}. Based on qualitative comparison, we select GPT-5 mini as the generator used
for all dataset construction. GPT-5 mini consistently produces step-by-step reasoning, which improves transparency and supports the use of the \textit{reasoning} field as a high-quality supervisory signal. Figure~\ref{fig:stage3_gpt5mini} illustrates a qualitative example of the generated comparative QA group for a single track pair. A broader qualitative comparison against other candidate generators is deferred to Appendix~\ref{sec:appendix-generators}.

\subsection{Stage 4: Evaluation and Quality-Based Filtering}
\label{sec:stage4}

After generating the full set of 39,291 comparative QA items in Stage~3,
we evaluate and refine the dataset through a final quality-control stage.
This stage consists of:  
(1) a two-phase evaluation (human evaluation and LLM-as-a-Judge), and  
(2) quality-based filtering to produce the final released dataset.

\paragraph{Human Evaluation.}
To obtain a human reference for evaluation reliability, we randomly sample
300 QA items from the GPT-5-mini generated dataset.
Four annotators independently rate each item using four criteria:
Correctness, Comparative Validity, Reasoning Quality, and Difficulty.

Table~\ref{tab:human_eval} reports per-annotator scores together with
aggregate statistics, including the human mean and mean absolute deviation (MAD),
for four evaluation criteria.
While individual item-level agreement is difficult to quantify due to
score compression on the 1--5 Likert scale, we observe consistent
\emph{mean-level trends} across annotators for the three semantic criteria
(Correctness, Comparative Validity, and Reasoning Quality).

Table~\ref{tab:human_eval} also includes scores from the GPT-5 Mini LLM judge
on the same subset.
Across all semantic criteria, LLM-based scores closely match human mean ratings,
indicating that the LLM judge can serve as a scalable proxy for semantic
quality assessment.

\begin{table*}[t]
\centering
\resizebox{\textwidth}{!}{%
\begin{tabular}{lcccc||ccc}
\toprule
\textbf{Metric} 
& \textbf{R1} & \textbf{R2} & \textbf{R3} & \textbf{R4}
& \textbf{Human Mean} 
& \textbf{GPT-5 Mini Mean} 
& \textbf{MAD} \\
\midrule
Correctness     
& 4.93 & 4.66 & 4.64 & 5.00 
& \textbf{4.79} 
& 4.87 
& 0.32 \\

Comparative Validity  
& 4.93 & 4.75 & 4.83 & 4.99 
& \textbf{4.83} 
& 4.61 
& 0.44 \\

Reasoning Quality        
& 4.87 & 4.71 & 4.67 & 5.00 
& \textbf{4.78} 
& 4.37 
& 0.56 \\

Difficulty       
& 1.89 & 2.58 & 3.12 & 2.66 
& \textbf{2.25} 
& 2.17 
& 0.57 \\
\bottomrule
\end{tabular}%
}
\caption{
Human evaluation results on 300 sampled QA items using four annotators (R1--R4),
together with LLM-as-a-Judge scores on the same subset.
We report human mean scores and mean absolute deviation (MAD) to characterize
aggregate-level agreement between human and LLM judgments.
}
\label{tab:human_eval}
\end{table*}

\paragraph{LLM-as-a-Judge Evaluation.}
Based on the observed alignment between human and LLM judgments on the
300-item subset, we employ GPT-5-mini as an LLM-as-a-Judge to evaluate
the full dataset at scale.
Rather than treating the LLM as a replacement for human evaluation,
we use it as a \emph{consistent semantic scorer} whose behavior is
anchored by human reference statistics. This approach enables scalable quality control while preserving comparability with human judgments at the level of aggregated trends. As shown in Table~\ref{tab:human_eval}, the LLM judge reproduces the relative ordering and mean scores of the semantic criteria, even though individual item-level agreement may vary.

\paragraph{Quality-Based Filtering.}
For the final release, we retain only QA groups for which all three QA items receive perfect scores (5/5/5) across the three semantic criteria (Correctness, Comparative Validity, and Reasoning Quality) under LLM-based evaluation.
This conservative filtering strategy prioritizes semantic reliability
over coverage and is intended to support high-precision benchmarking
and supervised training.

Importantly, this step does not aim to simplify the task itself.
As shown in Section~\ref{sec:stats}, the filtered dataset preserves
the original diversity of genres, topics, and cross-genre pairings.
The resulting dataset contains \textbf{12,173} track pairs
(\textbf{92.9\%} of the original 13,097 pairs), forming the final
released benchmark.


\subsection{Dataset Statistics}
\label{sec:stats}

Our final dataset contains \textbf{12,173} track pairs, each associated with
three comparative questions, resulting in a total of \textbf{36,519} QA items.
The dataset exhibits a strong dominance of cross-genre comparisons,
encouraging fine-grained comparative reasoning beyond genre identity.

Human difficulty ratings collected during Stage~4 indicate that
sentence-level questions are substantially more challenging than yes/no and
short-answer formats, reflecting their requirement for multi-attribute
comparison and coherent natural-language justification.

A detailed statistical analysis, including genre distributions,
genre-pair frequencies, question-topic coverage, and question-type difficulty,
is provided in Appendix~\ref{sec:appendix-stats} and
Figure~\ref{fig:genre_distribution}.

\section{Baseline Benchmark}

To assess the difficulty of Jamendo-MT-QA and provide reference points for
future work, we benchmark several representative audio--language models on
a sampled subset of the dataset.
Our goal is not to exhaustively optimize model performance, but to
characterize how existing models perform on multi-track comparative
question answering under a unified evaluation protocol.

\subsection{Baselines}

\begin{figure*}[t]
  \centering
  \includegraphics[width=\textwidth]{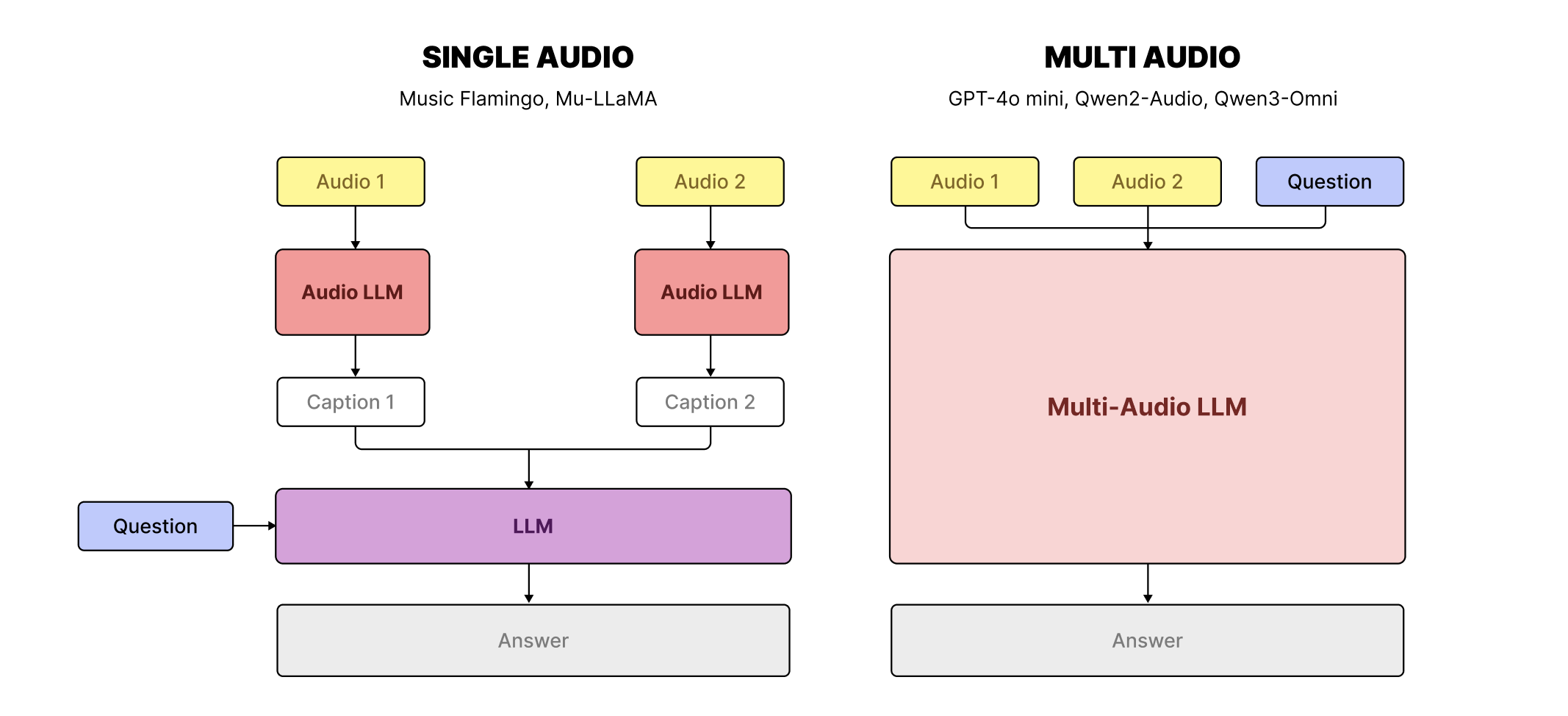}
  \caption{
  Baseline evaluation setups for Jamendo-MT-QA.
  \textbf{Left: caption-based single-audio baselines} (e.g., Music Flamingo, MU-LLaMA), where each track is independently captioned and an LLM performs comparative reasoning over the two captions plus the question.
  \textbf{Right: multi-audio baselines} (e.g., GPT-4o(-mini) Audio, Qwen2-Audio, Qwen3-Omni), which take two audio inputs and the question to produce an answer end-to-end.
  }
  \label{fig:baseline_setup}
\end{figure*}

Figure~\ref{fig:baseline_setup} summarizes our two baseline setups.
Some models can directly process multiple audio inputs (\textit{Multi}),
while others operate on single-audio inputs and are evaluated via an intermediate caption representation (\textit{Cap}).
This design helps disentangle limitations of multi-audio perception from those of comparative reasoning over high-level semantics.

We evaluate the following baseline models:
(1) \textbf{GPT-4o Audio} and \textbf{GPT-4o-mini Audio}\cite{gpt4}, which accept multiple audio inputs and perform end-to-end audio--text reasoning;
(2) \textbf{Qwen2-Audio} \cite{qwen2audio} and \textbf{Qwen3-Omni} \cite{qwenomni}, open-source multimodal models with multi audio input support; and
(3) \textbf{Music Flamingo}, a strong audio--language model,
evaluated in a caption-based comparative setting by comparing generated captions.
Due to cost constraints, GPT-4o Audio and GPT-4o-mini Audio are
evaluated on a randomly sampled subset of 2,010 track pairs,
while other baselines follow the same evaluation protocol when applicable.

\paragraph{Caption-based Single-Audio Baselines.}
Not all audio--language models natively support multiple audio inputs. To enable a broader and more diagnostic comparison, we include caption-based single-audio baselines, where each track is first independently converted into a natural language description and comparative reasoning is performed purely in the text domain by an LLM.

This setup allows us to disentangle the contribution of direct multi-audio perception from that of semantic abstraction and comparative reasoning. By evaluating caption-based and multi-audio models under a unified protocol, we assess whether performance gains primarily stem from access to multiple audio inputs or from higher-level reasoning over structured representations.

\subsection{Evaluation Metrics}

We evaluate three types of questions separately.
For \textit{yes/no} questions, we report accuracy.
For \textit{short-answer} questions, accuracy is computed by exact match
between the predicted track identifier and the ground-truth answer.
For \textit{sentence-level} questions, we report BLEU \cite{BLEU} and ROUGE-1,2,L \cite{ROUGE} as
surface-form similarity metrics based on n-gram overlap, and BERTScore \cite{BERTScore}
as a semantic similarity metric.

Since such metrics are known to be imperfect for open-ended comparative
generation, we additionally employ an LLM-as-a-Judge protocol \cite{llmasajudge} that assigns
a 1--5 score to sentence-level answers based on correctness and comparative
soundness.
Full details of the LLM-as-a-Judge prompt and scoring rubric used for baseline
evaluation are provided in Appendix~\ref{sec:appendix-judge}.

\subsection{Results}
\begin{table*}[t]
\centering
\small
\setlength{\tabcolsep}{6pt}
\begin{tabular}{l l c c c c c c c c c}
\toprule
\textbf{Model} & \textbf{Type}
& \multicolumn{1}{c}{\textbf{Yes/No}}
& \multicolumn{1}{c}{\textbf{Short}}
& \multicolumn{7}{c}{\textbf{Sentence-level}} \\
\cmidrule(lr){3-3}\cmidrule(lr){4-4}\cmidrule(lr){5-11}
& &
\textbf{Acc.} & \textbf{Acc.}
& \textbf{BLEU} & \textbf{R-1} & \textbf{R-2} & \textbf{R-L}
& \textbf{BERT-F1} & \textbf{GPT} & \textbf{Claude} \\
\midrule

Music Flamingo & Cap
& \underline{\textbf{77.4\%}} & \underline{\textbf{89.7\%}}
& \underline{\textbf{4.00}} & \textbf{27.5} & \textbf{4.6} & \underline{\textbf{24.7}}
& \underline{\textbf{0.8786}} & \underline{\textbf{3.24}} & \underline{\textbf{3.87}} \\

Qwen2-Audio & Cap
& 37.4\% & 39.1\%
& 1.88 & 19.3 & 1.8 & 17.4
& 0.8489 & 1.49 & 1.53 \\

MU-LLaMA & Cap
& 20.6\% & 55.3\%
& 2.39 & 23.8 & 3.2 & 16.5
& 0.8572 & 2.36 & 2.01 \\

\cmidrule(lr){1-11}

Qwen2-Audio & Multi
& 50.9\% & 80.2\%
& 2.09 & 21.0 & 3.4 & 14.6
& 0.8472 & 1.37 & 1.62 \\

Qwen3-Omni & Multi
& \textbf{62.9\%} & \textbf{80.3\%}
& \textbf{3.58} & \underline{\textbf{29.3}} & \underline{\textbf{6.6}} & \textbf{20.4}
& \textbf{0.8632} & \textbf{3.11} & \textbf{3.48} \\

\bottomrule
\end{tabular}
\caption{Results on the full benchmark of 12,173 track pairs.}
\label{tab:results_12173}
\end{table*}

\begin{table*}[t]
\centering
\small
\setlength{\tabcolsep}{6pt}
\begin{tabular}{l l c c c c c c c c c}
\toprule
\textbf{Model} & \textbf{Type}
& \multicolumn{1}{c}{\textbf{Yes/No}}
& \multicolumn{1}{c}{\textbf{Short}}
& \multicolumn{7}{c}{\textbf{Sentence-level}} \\
\cmidrule(lr){3-3}\cmidrule(lr){4-4}\cmidrule(lr){5-11}
& &
\textbf{Acc.} & \textbf{Acc.}
& \textbf{BLEU} & \textbf{R-1} & \textbf{R-2} & \textbf{R-L}
& \textbf{BERT-F1} & \textbf{GPT} & \textbf{Claude} \\
\midrule

Music Flamingo & Cap
& \underline{\textbf{82.1\%}} & \underline{\textbf{88.1\%}}
& \underline{\textbf{2.06}} & \underline{\textbf{25.3}} & \underline{\textbf{5.9}} & \underline{\textbf{19.6}}
& \underline{\textbf{0.875}} & \underline{\textbf{3.25}} & \underline{\textbf{3.97}} \\

Qwen2-Audio & Cap
& 26.9\% & 55.5\%
& 0.58 & 15.3 & 1.6 & 12.1
& 0.845 & 1.37 & 1.60 \\

MU-LLaMA & Cap
& 23.6\% & 50.5\%
& 0.68 & 16.1 & 1.9 & 13.3
& 0.853 & 1.92 & 2.13 \\

\cmidrule(lr){1-11}

Qwen2-Audio & Multi
& 34.4\% & 77.7\%
& 0.57 & 14.3 & 1.4 & 11.3
& 0.845 & 1.63 & 1.75 \\

Qwen3-Omni & Multi
& 59.7\% & 75.5\%
& \textbf{1.22} & 19.9 & \textbf{3.9} & 16.0
& 0.860 & 2.57 & 3.25 \\

GPT-4o-mini-Audio & Multi
& \textbf{77.3\%} & 73.2\%
& 0.87 & 17.8 & 2.7 & 14.7
& 0.859 & 2.92 & 3.14 \\

GPT-4o-Audio & Multi
& 69.6\% & \textbf{84.4\%}
& 1.04 & \textbf{20.0} & 3.6 & \textbf{16.7}
& \textbf{0.870} & \textbf{3.17} & \textbf{3.57} \\

\bottomrule
\end{tabular}
\caption{Results on the subset of 2,010 track pairs.}
\label{tab:results_2010}
\end{table*}

We evaluate baseline models on Jamendo-MT-QA under two evaluation settings,
summarized in Tables~\ref{tab:results_12173} and~\ref{tab:results_2010}.
To additionally assess models that support end-to-end multi-audio reasoning,
we report results on a smaller subset of \textbf{2,010} track pairs that can be processed
by all models, including GPT-based models.

Across models and evaluation settings, BLEU and ROUGE scores remain low,
reflecting the difficulty of generating precise surface-form matches for
open-ended comparative explanations.
In contrast, BERTScore and LLM-as-a-Judge scores are substantially higher,
indicating that models often produce semantically correct and comparative
answers despite limited lexical overlap.
This gap highlights the limitations of n-gram overlap metrics for evaluating
sentence-level questions.

A notable result is that the caption-based \textbf{Music Flamingo} baseline
achieves consistently strong performance across both evaluation settings.
As shown in Tables~\ref{tab:results_12173} and~\ref{tab:results_2010},
Music Flamingo attains the highest sentence-level BLEU, ROUGE, and
LLM-as-a-Judge scores among caption-based models, and remains competitive
with multi-audio systems on the subset of 2,010 track pairs.
Despite operating on single-audio inputs, Music Flamingo benefits from
high-quality captions that capture rich musical attributes, which
support comparative reasoning when combined with an LLM.

Multi-audio models show mixed performance on yes/no and short-answer questions.
In particular, \textbf{GPT-4o Audio} and \textbf{GPT-4o-mini Audio} achieve
strong accuracy on the subset of 2,010 track pairs, while open-source multi-audio
models exhibit more variable results.
However, gains on sentence-level questions remain modest across all
multi-audio models, suggesting that direct access to multiple audio inputs
alone does not guarantee strong comparative explanation quality.

Across all models, performance is consistently higher on yes/no and
short-answer questions than on sentence-level questions.
This indicates that identifying the correct comparative direction is often
easier than generating well-grounded natural language explanations
that articulate perceptual differences across tracks.
Sentence-level comparative reasoning therefore remains the primary bottleneck
for current audio--language models.

Overall, these results show that performance on Jamendo-MT-QA is not
determined solely by multi-audio perception.
Instead, the quality of intermediate semantic representations and a model’s
ability to reason over them play a critical role, particularly for
sentence-level comparative explanations.
This finding is consistent with prior work in multimodal learning, which shows
that strong abstraction and language alignment can partially compensate for
limited perceptual access in complex reasoning tasks
\cite{elizalde2023clap,mulan,musicflamingo,mullama}.

\paragraph{Key Observations.}
We draw three main conclusions from the baseline benchmark.

First, models consistently perform better on yes/no and short-answer questions
than on sentence-level questions, indicating that identifying comparative
direction is easier than generating explicit comparative explanations.

Second, caption-based single-audio models are not uniformly inferior to
multi-audio models, highlighting the importance of representation quality and
reasoning capability in comparative Music-QA.

Third, the large discrepancy between surface-form automatic metrics
(BLEU/ROUGE) and LLM-as-a-Judge scores underscores the inadequacy of n-gram
overlap metrics for evaluating comparative music explanations, motivating the
use of semantically grounded evaluation protocols.

\subsection{Error Analysis}
\label{sec:error_analysis}

\begin{table}[t]
\centering
\small
\begin{tabular}{l|ccc}
\toprule
\textbf{Model} 
& \textbf{Comp.Col.} 
& \textbf{Attr.Hal.} 
& \textbf{Gran.Mis.} \\
\midrule
Music Flamingo 
& 35.3\% & 64.0\% & 0.7\% \\

MU-LLaMA 
& 37.3\% & 60.7\% & 2.0\% \\

GPT-4o-Audio 
& 56.7\% & 35.3\% & 8.0\% \\

Qwen3-Omni 
& 23.4\% & 75.9\% & 0.7\% \\
\bottomrule
\end{tabular}
\caption{
Error type distribution across models on sentence-level questions
(LLM judge score $<$ 3). Classified by GPT-4o-mini.
}
\label{tab:error_analysis}
\end{table}

\begin{figure}[t]
  \centering
  \includegraphics[width=\columnwidth]{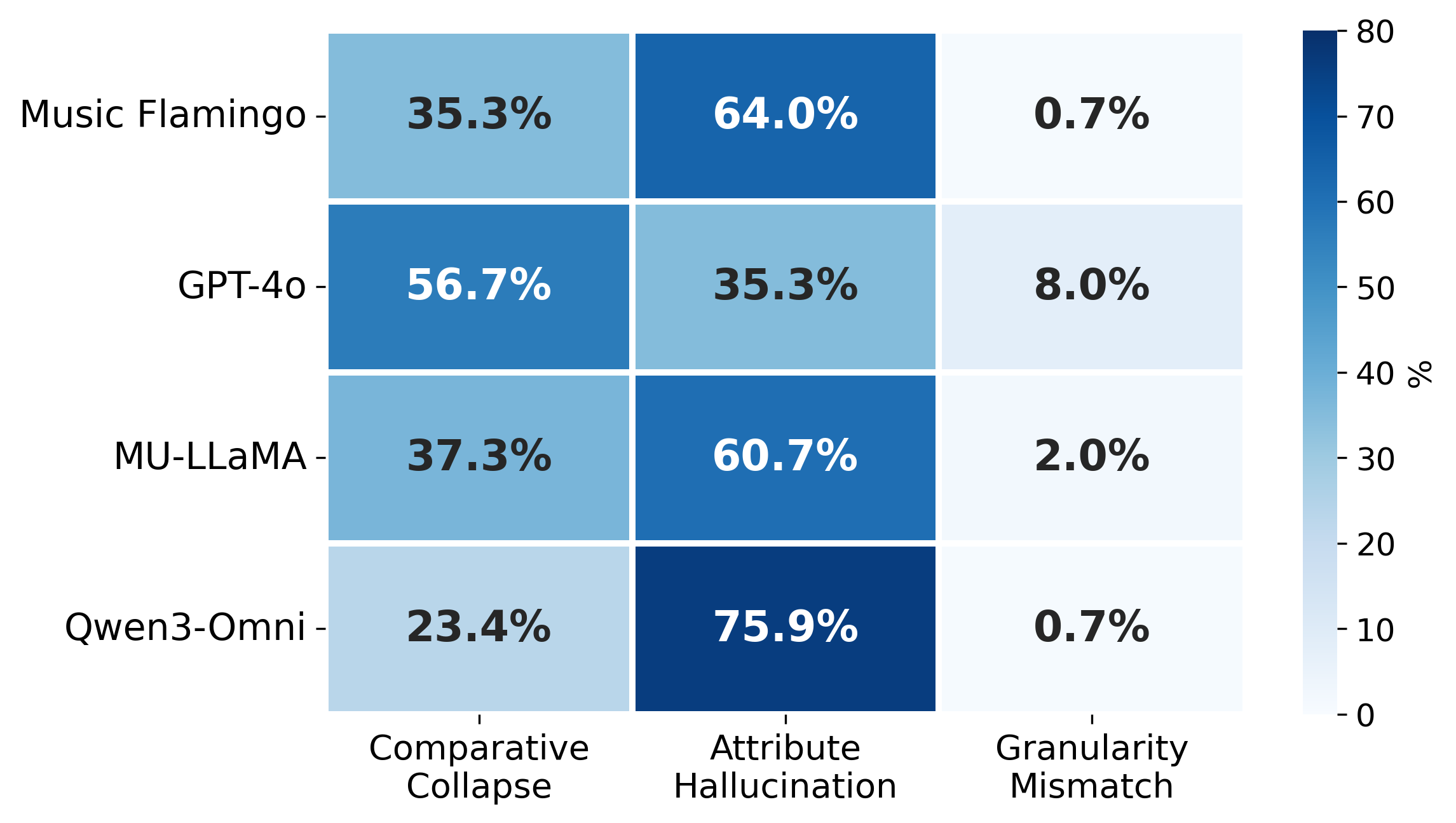}
  \caption{
  Heatmap visualization of error type distributions across models.
  Values indicate the percentage of each error category among incorrectly
  answered sentence-level questions (LLM judge score $<$ 3).
  }
  \label{fig:error_heatmap}
\end{figure}

To better understand the failure modes of current audio--language models,
we conduct an error analysis on incorrectly answered \emph{sentence-level}
questions, combining qualitative categorization with a quantitative breakdown of dominant error types. Table~\ref{tab:error_analysis} and Figure~\ref{fig:error_heatmap} summarize the
distribution of dominant error types across models, computed over incorrectly
answered sentence-level questions.

\paragraph{Error Categories.}
We define the following error types:
\textbf{(A) Comparative Collapse}, where the model avoids explicit comparison
and produces a generic summary of both tracks;
\textbf{(B) Attribute Hallucination}, where the model introduces musical
attributes not supported by the audio or captions;
\textbf{(C) Granularity Mismatch}, where the comparison is too coarse or too
fine-grained relative to the question intent.

\paragraph{Model-Specific Failure Patterns.}
We observe clear qualitative differences across models (Table~\ref{tab:error_analysis};
Figure~\ref{fig:error_heatmap}).
\textbf{Music Flamingo} primarily fails through \emph{attribute hallucination},
suggesting that while the comparative structure is often preserved,
factual grounding of fine-grained musical attributes remains challenging in
caption-based settings.
\textbf{MU-LLaMA} shows a similar pattern, but with a slightly higher rate of
\emph{comparative collapse}, indicating weaker comparative commitment even when
high-level content is partially captured.
In contrast, \textbf{GPT-4o-Audio} exhibits the highest rate of
\emph{comparative collapse}, frequently producing fluent but non-committal
summaries that avoid explicit relational judgments.
Finally, \textbf{Qwen3-Omni} is dominated by \emph{attribute hallucination},
often introducing unsupported instrument or production details despite making
explicit comparisons. Representative examples for each model illustrating these failure modes are
provided in Appendix~\ref{sec:appendix-error}.

Overall, these results highlight that Jamendo-MT-QA does not merely measure
accuracy, but also exposes \emph{qualitatively distinct failure modes} in
multi-track music reasoning, underscoring its value as a diagnostic benchmark.

\section{Conclusion}
\label{sec:conclusion}

We introduced Jamendo-MT-QA, a multi-track comparative question answering dataset built on top of Jamendo-QA, targeting Music-QA systems that must reason about relationships between tracks rather than isolated audio clips.
Our LLM-assisted pipeline generates three comparative questions per track pair, covering yes/no, short-answer, and sentence-level reasoning.

A key aspect of our dataset construction is the integration of human evaluation and LLM-as-a-Judge evaluation.
By validating strong alignment between human judgments and LLM-based scores on a representative subset, we enable scalable quality control while maintaining consistency with human semantic evaluation.
This approach allows us to filter and curate a large dataset without sacrificing annotation reliability.

In addition, we benchmark representative audio--language models on Jamendo-MT-QA.
While current models achieve reasonable performance on yes/no and short-answer comparisons, sentence-level comparative generation remains challenging, especially under surface-form automatic metrics.

We believe Jamendo-MT-QA provides a valuable resource for future research on comparative music understanding, multi-track reasoning, and audio-language modeling.

\section*{Limitations}
Our benchmark relies on LLM-generated comparative QA, which may inherit
generator-specific biases despite human and judge-based filtering.
Automatic metrics (BLEU/ROUGE) are limited for sentence-level comparative
answers, and LLM-as-a-Judge introduces model-dependent scoring noise.
Finally, some baselines require caption-based evaluation due to lack of multi-audio support, which may underestimate the potential of
end-to-end audio reasoning models.

\section*{Ethical Considerations}

Our dataset is derived from Creative Commons-licensed Jamendo audio, and we respect the original licensing conditions by releasing only annotations and metadata.
While question generation is automated, we monitor and filter for potentially sensitive or biased content.
Nevertheless, the dataset may reflect existing biases in genre labels, artist demographics, and descriptive language.
We encourage practitioners to consider these factors when training or evaluating models on this resource.

\paragraph{Human Subjects.}
The human annotation study was conducted for research quality control
on synthetic QA data and did not involve the collection of personal or
sensitive information.
Annotators provided informed consent, and the study was determined to be
exempt from formal IRB review under institutional guidelines.

\section*{Acknowledgments}
This research was supported by the Basic Science Research Program through the National Research Foundation of Korea (NRF) funded by the Ministry of Education (RS-2025-25422688), 
and an IITP grant funded by the Korean Government (MSIT) (No. RS-2020-II201361, Artificial Intelligence Graduate School Program, Yonsei University).


\bibliography{custom}

\appendix
\section{Human and LLM Evaluation Criteria}
\label{sec:appendix-eval}

This appendix provides the full descriptions of the five evaluation criteria
used in the human evaluation and the four criteria used in the LLM-as-a-Judge
evaluation.

\subsection{Human Evaluation Criteria}
\label{sec:human-evaluation}

We conduct human evaluation in two separate stages. In \textbf{Stage 1}, we verify
the \textit{audio--text alignment} of the enriched captions generated by Music Flamingo,
where annotators have access to the underlying audio. In \textbf{Stage 4}, we use a
multi-criteria rubric to evaluate and filter comparative QA items based on the
provided text descriptions of Track~A and Track~B.

\begin{table}[ht]
\centering
\small
\setlength{\tabcolsep}{5pt}
\begin{tabular}{lcc}
\toprule
\textbf{Annotator} & \textbf{Mean} & \textbf{Std} \\
\midrule
Annotator A & 4.43 & 0.65 \\
Annotator B & 4.47 & 0.62 \\
Annotator C & 4.60 & 0.71 \\
Annotator D & 4.85 & 0.48 \\
\midrule
\textbf{Average} & \textbf{4.59} & -- \\
\bottomrule
\end{tabular}
\caption{
Human evaluation results for audio--text alignment of enriched captions
in Stage~1, reported on a 1--5 scale.
}
\label{tab:stage1_audio_text_alignment}
\end{table}

\paragraph{Stage 1: Audio-Text Alignment of Enriched Captions (Music Flamingo)}
To ensure that the enriched captions faithfully describe the actual music content,
four human annotators listen to the audio and rate the caption-level alignment on a
\textbf{1--5 scale}:
\textit{“Does the enriched caption accurately align with the musical content of the audio
(e.g., instrumentation, vocals, tempo, mood, and production cues)?”}
Scores range from \textbf{1 (complete mismatch)} to \textbf{5 (perfect alignment)},
with intermediate values (2--4) reflecting partial correctness and varying degrees of
coverage and specificity. Given the consistently high alignment scores and low variance across annotators,
we found that this evaluation size was sufficient to validate caption quality
for downstream QA generation.

Table~\ref{tab:stage1_audio_text_alignment} summarizes the results of the
human evaluation for audio--text alignment of enriched captions.
Across 1200 individual ratings, the captions achieve a high overall
average score of \textbf{4.59}, indicating strong consistency between
the generated textual descriptions and the underlying music audio.
The equal-weight average across annotators (4.59) further confirms that
this alignment is consistent and not dominated by a single evaluator.
These results suggest that the enriched captions provide a reliable
semantic foundation for subsequent QA generation stages.

\paragraph{Stage 4: Comparative QA Evaluation and Filtering Criteria}
For the evaluation and filtering in Stage 4, annotators rate each model answer
(or candidate QA item, depending on the filtering stage) on a \textbf{1--5 scale}
according to the following dimensions, grounded in the provided textual descriptions
of Track~A and Track~B (i.e., captions/metadata, without requiring direct audio access):

\textbf{(1) Correctness.}
\textit{“Is the answer factually correct based on the provided descriptions of Track~A and Track~B?”}
Scores range from \textbf{1 (completely incorrect)} to \textbf{5 (completely correct)},
with 2--4 indicating partially correct answers (e.g., correct direction but missing key evidence).

\textbf{(2) Comparative Validity.}
\textit{“Does the answer make a valid comparative statement between Track~A and Track~B?”}
Scores range from \textbf{1 (no comparison or invalid comparison)} to \textbf{5 (excellent comparative statement)},
with 2--4 capturing weak, ambiguous, or only partially comparative responses.

\textbf{(3) Reasoning Quality.}
\textit{“Is the reasoning coherent, logically grounded in the provided descriptions, and properly supporting the answer?”}
Scores range from \textbf{1 (incoherent/illogical)} to \textbf{5 (excellent, well-supported reasoning)},
with intermediate values reflecting partially grounded reasoning or missing justification.

\textbf{(4) Difficulty.}
\textit{“How difficult is the question to answer correctly?”}
Scores range from \textbf{1 (very easy)} to \textbf{5 (very difficult)},
where higher difficulty indicates that answering requires subtle distinctions,
multi-attribute comparison, or non-trivial inference from the descriptions.

\subsection{Caption Attribute Coverage}
\label{sec:appendix-coverage}

We conducted a systematic attribute coverage analysis on all 7,335 Music Flamingo captions to verify whether captions consistently include the seven core musical attributes. Table~\ref{tab:caption_coverage} shows that every individual attribute exceeds 90\% coverage, and 85.6\% of captions contain all seven attributes simultaneously, confirming that the caption generation prompt was effective.

\begin{table}[ht]
\centering
\small
\begin{tabular}{lcc}
\toprule
\textbf{Musical Attribute} & \textbf{Coverage} & \textbf{Count} \\
\midrule
Instrumentation & 99.9\% & 7,325 / 7,335 \\
Production style & 99.7\% & 7,316 / 7,335 \\
Vocal characteristics & 99.7\% & 7,316 / 7,335 \\
Mood / atmosphere & 98.4\% & 7,216 / 7,335 \\
Tempo & 96.7\% & 7,095 / 7,335 \\
Genre & 96.3\% & 7,061 / 7,335 \\
Key / tonality & 90.8\% & 6,661 / 7,335 \\
\midrule
All 7 attributes & 85.6\% & 6,280 / 7,335 \\
\bottomrule
\end{tabular}
\caption{Caption attribute coverage across all 7,335 Music Flamingo captions.}
\label{tab:caption_coverage}
\end{table}

\subsection{LLM-as-a-Judge Criteria}

The automatic evaluation follows the same rubric as the human evaluation,
with the exception of the Audio-Text Alignment, which cannot be reliably
assessed by text-only LLM judges without access to the underlying audio.

Accordingly, LLM evaluators score each answer using the following four dimensions:
\textbf{Correctness}, \textbf{Comparative Validity}, \textbf{Reasoning Quality},
and \textbf{Difficulty}.  
Each criterion is rated on a \textbf{1--5 scale}, where intermediate scores
(2--4) represent partial correctness, weak or ambiguous comparisons, or varying
degrees of reasoning quality and difficulty.

\section{Output Schema for Comparative QA Generation}
\label{sec:output-schema}

The following listing shows the JSON output format used during Stage~3
comparative QA generation. For each track pair, the model generates three comparative question--answer pairs: yes/no, short-answer, and sentence-level.

\lstdefinestyle{jsonstyle}{
  basicstyle=\ttfamily\footnotesize,
  breaklines=true,
  frame=single,
  columns=fullflexible,
  showstringspaces=false,
  tabsize=2
}

\begin{lstlisting}[style=jsonstyle]
{
  "audio1": "{audio1}",
  "audio2": "{audio2}",
  "qa_pairs": [
    {
      "type": "yes_no",
      "question": "...",
      "reasoning": "...",
      "answer": "yes or no"
    },
    {
      "type": "short_answer",
      "question": "...",
      "reasoning": "...",
      "answer": "{audio1} or {audio2}"
    },
    {
      "type": "sentence",
      "question": "...",
      "reasoning": "...",
      "answer": "complete comparative sentence"
    }
  ]
}
\end{lstlisting}

\section{Prompt Template for Multi-Track QA Generation}
\label{sec:prompt-template}

Listing~\ref{lst:multiqa_prompt} shows the prompt template used for
multi-track comparative QA generation in Stage~3.
The prompt explicitly enforces comparative constraints and a fixed
three-question structure (yes/no, short-answer, and sentence-level)
for each audio pair.

\lstdefinestyle{promptstyle}{
  basicstyle=\ttfamily\footnotesize,
  breaklines=true,
  frame=single,
  columns=fullflexible,
  showstringspaces=false
}

\begin{lstlisting}[style=promptstyle, caption={Prompt template for multi-track comparative QA generation.}, label={lst:multiqa_prompt}]
You are a music comparison expert. Given two music tracks in JSON format,
you must generate exactly 3 COMPARATIVE question-answer pairs that directly
compare both tracks.

IMPORTANT: Every question MUST compare both tracks.
Do NOT ask about a single track alone.

Music Track 1:
{music1_str}

Music Track 2:
{music2_str}

Requirements:
1. YES/NO question - Must compare an attribute between both tracks
2. short-answer question - Answer should be the audio name
3. SENTENCE question - Detailed comparison in a complete sentence

\end{lstlisting}

\section{LLM-as-a-Judge Prompt and Scoring Rubric}
\label{sec:appendix-judge}

\subsection{Judge Prompt for Sentence-level Evaluation}

For sentence-level comparative questions, we employ an LLM-as-a-Judge protocol
to evaluate semantic correctness and comparative soundness.
The following prompt is used to score model predictions on a 0--5 scale.

\lstdefinestyle{judgestyle}{
  basicstyle=\ttfamily\footnotesize,
  breaklines=true,
  frame=single,
  columns=fullflexible,
  showstringspaces=false
}

\begin{lstlisting}[style=judgestyle]
You are an expert evaluator for music QA.
Compare the prediction with the ground truth answer.

## Question
{question}

## Ground Truth Answer
{ground_truth}

## Model Prediction
{prediction}

## Scoring Criteria (0-5 scale)
- 0: Completely wrong or irrelevant
- 1: Mostly wrong with very minor correct elements
- 2: Partially correct but missing key points
- 3: Correct on main points but missing details or has minor errors
- 4: Mostly correct with only minor omissions
- 5: Fully correct, captures all key information

Evaluate semantic similarity and correctness.
The prediction does not need to use exact wording.

Focus on:
- genre comparison
- tempo or energy
- vocal characteristics
- production style
- mood

Respond with ONLY a JSON object:
{"score": <0-5>, "explanation": "<brief reason>"}
\end{lstlisting}
\FloatBarrier

\subsection{Scoring Rubric}
For sentence-level comparative questions, we evaluate model outputs using a
graded semantic scoring scheme that captures both correctness and comparative
soundness beyond surface-form overlap.
Table~\ref{tab:judge_rubric} summarizes the 0--5 rubric used by LLM judges.

\begin{table}[t]
\centering
\small
\resizebox{\columnwidth}{!}{%
\begin{tabular}{cl}
\toprule
\textbf{Score} & \textbf{Description} \\
\midrule
0 & Completely wrong or irrelevant \\
1 & Mostly wrong with very minor correct elements \\
2 & Partially correct but missing key points \\
3 & Correct on main points but missing details or minor errors \\
4 & Mostly correct with only minor omissions \\
5 & Fully correct, capturing all key comparative information \\
\bottomrule
\end{tabular}
}
\caption{
Scoring rubric used by LLM judges for sentence-level comparative answers.
}
\label{tab:judge_rubric}
\end{table}

\subsection{Cross-Model Self-Preference Bias}
\label{sec:appendix-bias}

To verify that the LLM evaluator does not exhibit self-preference bias, we analyzed all four evaluator--generator combinations. Table~\ref{tab:self_pref} shows that the GPT evaluator exhibits no statistically significant self-preference (all $p > 0.26$, Cohen's $d < 0.01$), and actually gives slightly higher scores to Claude-generated data.

\begin{table}[ht]
\centering
\small
\resizebox{\columnwidth}{!}{
\begin{tabular}{lccc}
\toprule
\textbf{Metric} & \textbf{GPT$\to$GPT} & \textbf{GPT$\to$Claude} & \textbf{$p$} \\
\midrule
Correctness & 4.729 & 4.741 & 0.318 \\
Comp.\ Validity & 4.721 & 4.735 & 0.264 \\
Reasoning Qual. & 4.721 & 4.735 & 0.264 \\
\bottomrule
\end{tabular}
}
\caption{Cross-model self-preference analysis (GPT evaluator, $N {\approx} 13{,}096$). No significant bias detected ($d < 0.01$).}
\label{tab:self_pref}
\end{table}

\subsection{Evaluation by Question Type}

Different question types are evaluated using task-appropriate strategies:

\begin{table}[t]
\centering
\small
\resizebox{\columnwidth}{!}{%
\begin{tabular}{lcc}
\toprule
\textbf{Question Type} & \textbf{Evaluation Method} & \textbf{Score Range} \\
\midrule
Yes/No & Rule-based matching & 0 or 1 \\
Short-answer & Rule-based matching (Track A/B) & 0 or 1 \\
Sentence-level & LLM-as-a-Judge & 0--5 \\
Open-ended & LLM-as-a-Judge & 0--5 \\
\bottomrule
\end{tabular}
}
\caption{
Evaluation methods used for different question types.
}
\label{tab:eval_by_type}
\end{table}

\section{Statistical Analysis of the Dataset}
\label{sec:appendix-stats}

This section provides a statistical analysis of the Jamendo
Multi-Track comparative QA dataset, focusing on genre--genre pair
composition, track-level genre distribution, same-versus-different genre
comparisons, and question topic diversity.

\subsection{Full Dataset Quality Verification}
\label{sec:appendix-fullverify}

We conducted automated factual verification across the entire dataset (12,173 pairs / 36,519 QAs). Table~\ref{tab:full_verify} summarizes the results. Every QA pair passes structural format validation, and all verifiable answer-level factual claims (tempo comparison, key identification) achieve 100\% consistency with source captions.

\begin{table}[ht]
\centering
\small
\begin{tabular}{lcc}
\toprule
\textbf{Verification Check} & \textbf{Tested} & \textbf{Pass} \\
\midrule
Answer format validity (non-empty) & 36,519 & 100.0\% \\
Yes/No: valid ``yes''/``no'' answer & 12,173 & 100.0\% \\
Short Answer: contains track name & 12,173 & 100.0\% \\
Sentence: answer $\geq$ 20 characters & 12,173 & 100.0\% \\
All 3 QA types present per pair & 12,173 & 100.0\% \\
Tempo answer correct (SA) & 1,627 & 100.0\% \\
Tempo answer correct (YN) & 1,586 & 100.0\% \\
Key answer correct (SA) & 3,673 & 100.0\% \\
Key answer correct (YN) & 306 & 100.0\% \\
\bottomrule
\end{tabular}
\caption{Full dataset quality verification (12,173 pairs).}
\label{tab:full_verify}
\end{table}

\subsection{Basic Statistics}

Table~\ref{tab:basic_stats} summarizes the fundamental properties of the dataset.
Each audio pair is associated with exactly three comparative QA items,
resulting in a total of 36,519 questions.

\begin{table}[ht]
\centering
\small
\begin{tabular}{lc}
\toprule
\textbf{Statistic} & \textbf{Value} \\
\midrule
Total music pairs & 12,173 \\
Total QA pairs & 36,519 \\
Unique genre pairs & 789 \\
\bottomrule
\end{tabular}
\caption{Basic statistics of the multi-track comparative QA dataset.}
\label{tab:basic_stats}
\end{table}

\subsection{Track-Level Genre Distribution}

Table~\ref{tab:genre_dist} presents the most frequent genres across all tracks
participating in pairwise comparisons.
The distribution is dominated by \textit{rock}, \textit{indie}, and \textit{pop},
which is consistent with the characteristics of Creative Commons music
collections such as Jamendo.

\begin{table}[ht]
\centering
\small
\begin{tabular}{lcc}
\toprule
\textbf{Genre} & \textbf{Count} & \textbf{Ratio} \\
\midrule
rock & 4,890 & 20.09\% \\
indie & 3,424 & 14.06\% \\
pop & 2,527 & 10.38\% \\
folk & 1,611 & 6.62\% \\
electronic & 1,446 & 5.94\% \\
rap & 1,358 & 5.58\% \\
metal & 1,072 & 4.40\% \\
rnb & 655 & 2.69\% \\
dance & 515 & 2.12\% \\
guitar & 479 & 1.97\% \\
\bottomrule
\end{tabular}
\caption{Top genres across all tracks in the dataset.}
\label{tab:genre_dist}
\end{table}

\subsection{Genre Pair Combinations}

\begin{figure*}[t]
\centering
\includegraphics[width=\textwidth]{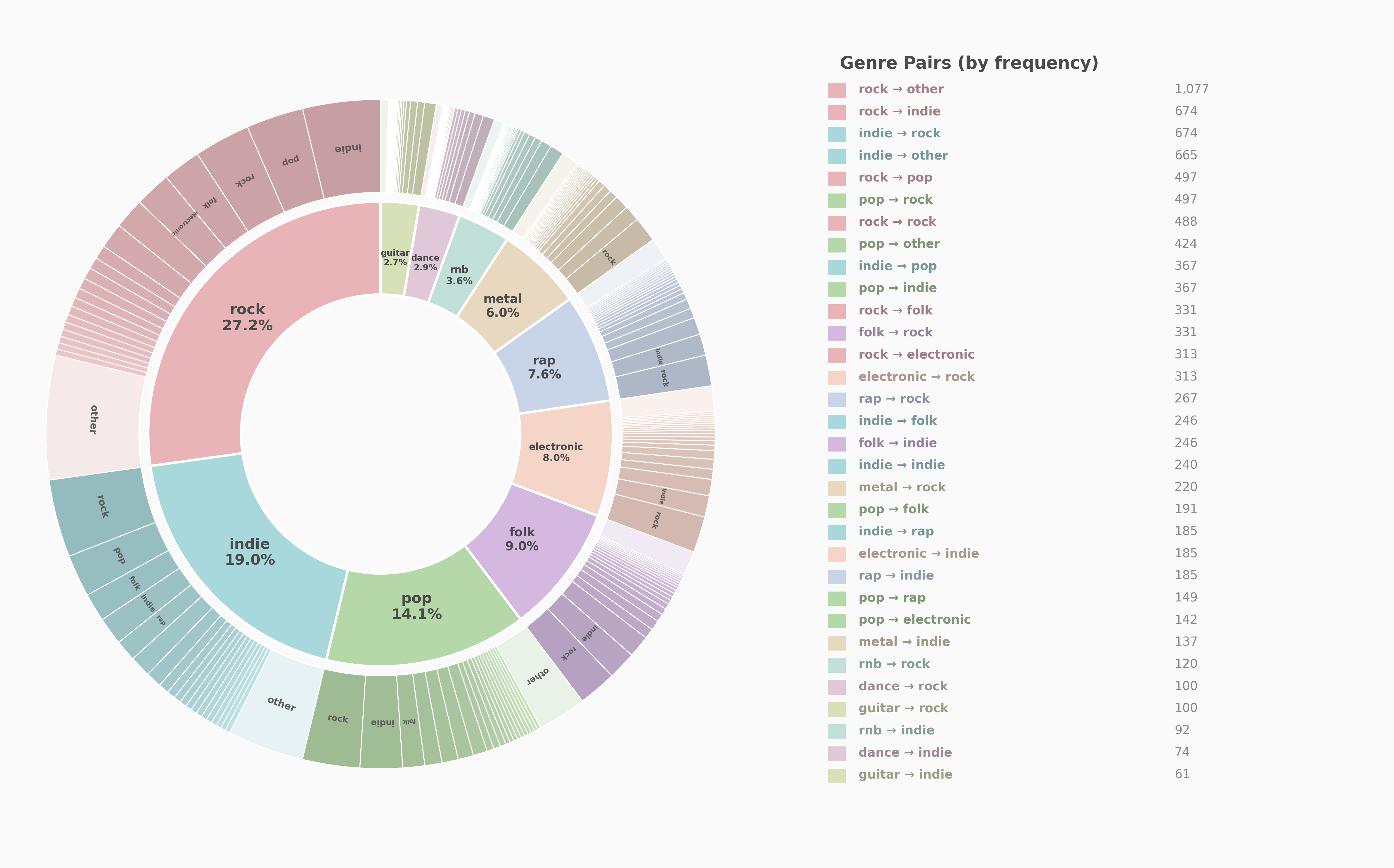}
\caption{
Genre composition and genre-pair distribution in Jamendo-MT-QA.
The inner ring shows the marginal distribution of individual genres across all tracks,
while the outer ring visualizes the frequency of genre--genre pairings used for
comparative question generation.
For clarity, only the top-10 most frequent genres are shown in the visualization;
therefore, percentages may differ from table-based statistics computed over all tracks.
}
\label{fig:genre_distribution}
\end{figure*}

Table~\ref{tab:genre_pairs} lists the most frequent genre--genre combinations
among the 12,173 track pairs.
While same-genre comparisons such as \textit{rock--rock} and
\textit{indie--indie} appear regularly, most pairs span different genres,
encouraging cross-style comparative reasoning.

Figure~\ref{fig:genre_distribution} provides a complementary visualization of
the genre structure in Jamendo-MT-QA.
The inner ring illustrates the marginal genre distribution of individual tracks,
confirming that the dataset spans a broad range of popular music styles,
including rock, indie, pop, folk, electronic, rap, and metal.

The outer ring depicts the distribution of genre--genre pairs used for
comparative question construction.
Consistent with the statistics reported in Table~\ref{tab:genre_pairs},
the majority of comparisons involve cross-genre pairs rather than same-genre
pairs.
This design choice encourages models to rely on fine-grained musical attributes
(e.g., instrumentation, production, tempo, or vocal characteristics) rather than
genre identity alone.

Overall, this visualization highlights the structural diversity of the dataset
and supports the claim that Jamendo-MT-QA systematically targets
\emph{cross-genre comparative reasoning} rather than trivial within-genre
comparisons.

\begin{table}[ht]
\centering
\small
\begin{tabular}{lcc}
\toprule
\textbf{Genre 1} & \textbf{Genre 2} & \textbf{Count} \\
\midrule
indie & rock & 674 \\
pop & rock & 497 \\
rock & rock & 488 \\
indie & pop & 367 \\
folk & rock & 331 \\
electronic & rock & 313 \\
rap & rock & 267 \\
folk & indie & 246 \\
indie & indie & 240 \\
metal & rock & 220 \\
\bottomrule
\end{tabular}
\caption{Most frequent genre--genre pairs in the dataset.}
\label{tab:genre_pairs}
\end{table}

\subsection{Same vs.\ Different Genre Pairs}

Table~\ref{tab:same_diff_genre} quantifies the proportion of same-genre and
different-genre track pairs.
A large majority of pairs (over 91\%) involve different genres, ensuring that
models must rely on fine-grained musical attributes rather than genre identity
alone.

\begin{table}[ht]
\centering
\small
\begin{tabular}{lcc}
\toprule
\textbf{Category} & \textbf{Count} & \textbf{Ratio} \\
\midrule
Same-genre pairs & 1,065 & 8.75\% \\
Different-genre pairs & 11,108 & 91.25\% \\
\bottomrule
\end{tabular}
\caption{Distribution of same-genre and different-genre track pairs.}
\label{tab:same_diff_genre}
\end{table}

\subsection{Question Topic Distribution}

Table~\ref{tab:topic_dist} summarizes the semantic topics addressed by the
comparative questions.
The dataset emphasizes perceptually salient musical attributes such as vocals,
instrumentation, production, and tempo, while also covering higher-level concepts
including key, mood, genre, and structural properties.

\begin{table}[ht]
\centering
\small
\begin{tabular}{lcc}
\toprule
\textbf{Topic} & \textbf{Count} & \textbf{Ratio} \\
\midrule
vocal & 15,509 & 42.47\% \\
tempo & 13,509 & 36.99\% \\
instrument & 13,384 & 36.65\% \\
production & 12,301 & 33.68\% \\
key & 11,086 & 30.36\% \\
mood & 4,646 & 12.72\% \\
genre & 3,203 & 8.77\% \\
structure & 594 & 1.63\% \\
\bottomrule
\end{tabular}
\caption{Distribution of question topics across the dataset.}
\label{tab:topic_dist}
\end{table}

Overall, the analysis confirms that the dataset exhibits substantial diversity
in both genre combinations and question semantics.
The dominance of cross-genre pairs (91.25\%) and the wide coverage of musical
attributes highlight the dataset’s suitability for evaluating multi-track
comparative reasoning in music understanding.

\subsection{Question-Type Difficulty}
\label{sec:appendix-difficulty}

\begin{table}[t]
\centering
\small
\begin{tabular}{lcc}
\toprule
\textbf{Question Type} & \textbf{Count} & \textbf{Avg.\ Difficulty} \\
\midrule
Yes/No & 12,173 & 2.5 \\
Short-Answer & 12,173 & 2.5 \\
Sentence-Level & 12,173 & 2.9 \\
\midrule
Total & 36,519 & 2.6 \\
\bottomrule
\end{tabular}
\caption{
Distribution of question types and their average difficulty scores from
Stage~4 human evaluation.
Difficulty is reported on a 1--5 scale, where higher values indicate more
challenging comparative reasoning.
}
\label{tab:question_type_difficulty}
\end{table}

To provide additional context on the relative hardness of different QA formats,
we report human-rated difficulty statistics for each question type.
Difficulty is rated on a 1--5 Likert scale during Stage~4 evaluation, where higher
scores indicate that answering requires more subtle cross-track distinctions,
multi-attribute comparison, or non-trivial inference from the provided descriptions.
As shown in Table~\ref{tab:question_type_difficulty}, sentence-level questions are
substantially more difficult than yes/no and short-answer questions, supporting our
claim that open-ended comparative explanation is the primary bottleneck for current models.

\section{Quantitative Error Distribution}
\label{sec:appendix-error}

This section provides implementation details for the quantitative error analysis reported in Section~\ref{sec:error_analysis}.

\paragraph{Scope.}
We analyze \emph{incorrectly answered} sentence-level questions, defined as items
whose LLM-as-a-Judge score is below 3.
For each model, percentages are computed by normalizing over the number of
incorrect sentence-level items for that model, so the error-type proportions sum
to 100\%.

\paragraph{Primary Error Labeling.}
Each incorrect item is assigned a \emph{primary} error type among the four
categories in Section~\ref{sec:error_analysis}.
We use GPT-4o-mini to classify the dominant failure mode based on the question,
reference answer, and the model's prediction.
When multiple issues co-occur (e.g., collapse \emph{and} hallucination), the
classifier is instructed to select the most salient error that explains the final answer's failure.

\subsection{Representative Error Examples}
Table~\ref{tab:error_examples} provides representative failure cases for each model,
illustrating typical breakdown patterns across the four error categories.

\section{Qualitative Comparison of Stage~3 Generators}
\label{sec:appendix-generators}

\begin{figure*}[t]
\centering
\includegraphics[width=\textwidth]{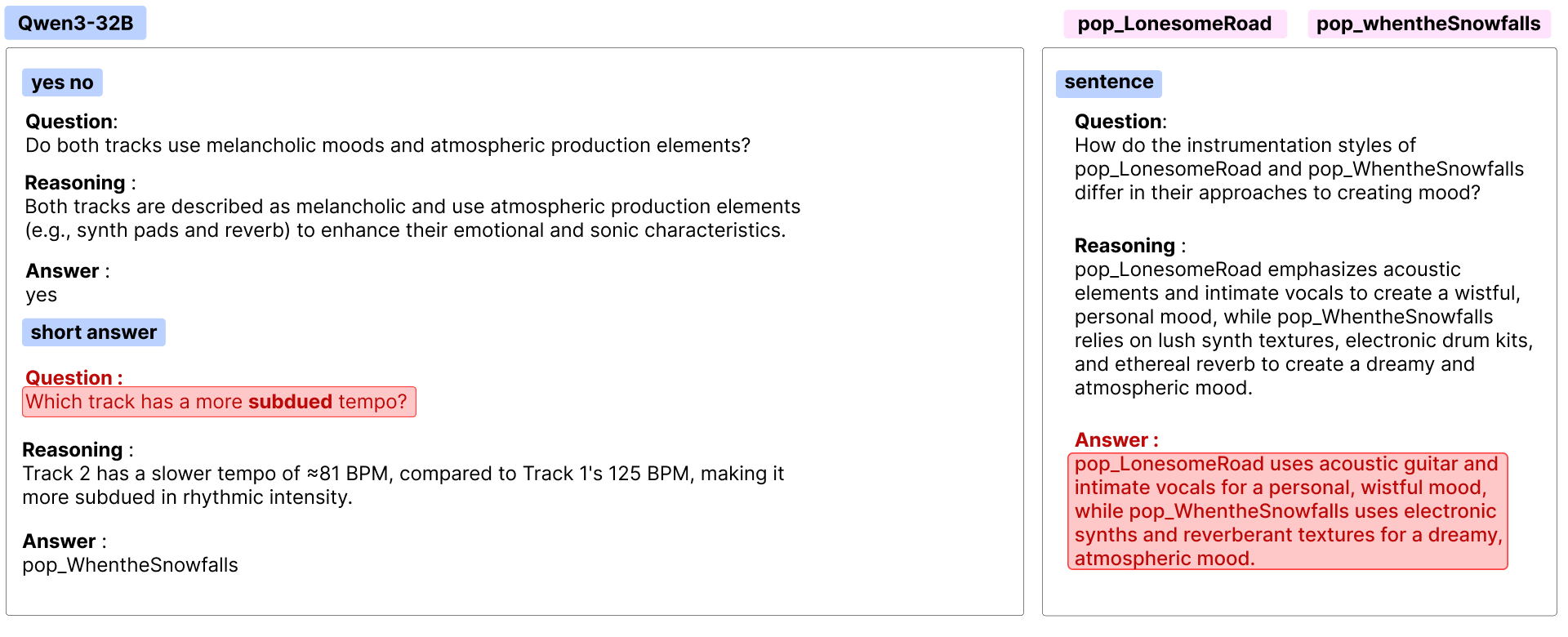}
\caption{
Qualitative example of Stage~3 comparative QA generation using \textbf{Qwen3-32B}.
The short-answer questions occasionally contain uncommon or subjective word choices,
and sentence-level answers often closely mirror the reasoning text.
Red highlights mark cases where reasoning and answers are nearly identical,
reducing diversity and usefulness for training.
}
\label{fig:stage3_qwen}
\end{figure*}

\begin{figure*}[t]
\centering
\includegraphics[width=\textwidth]{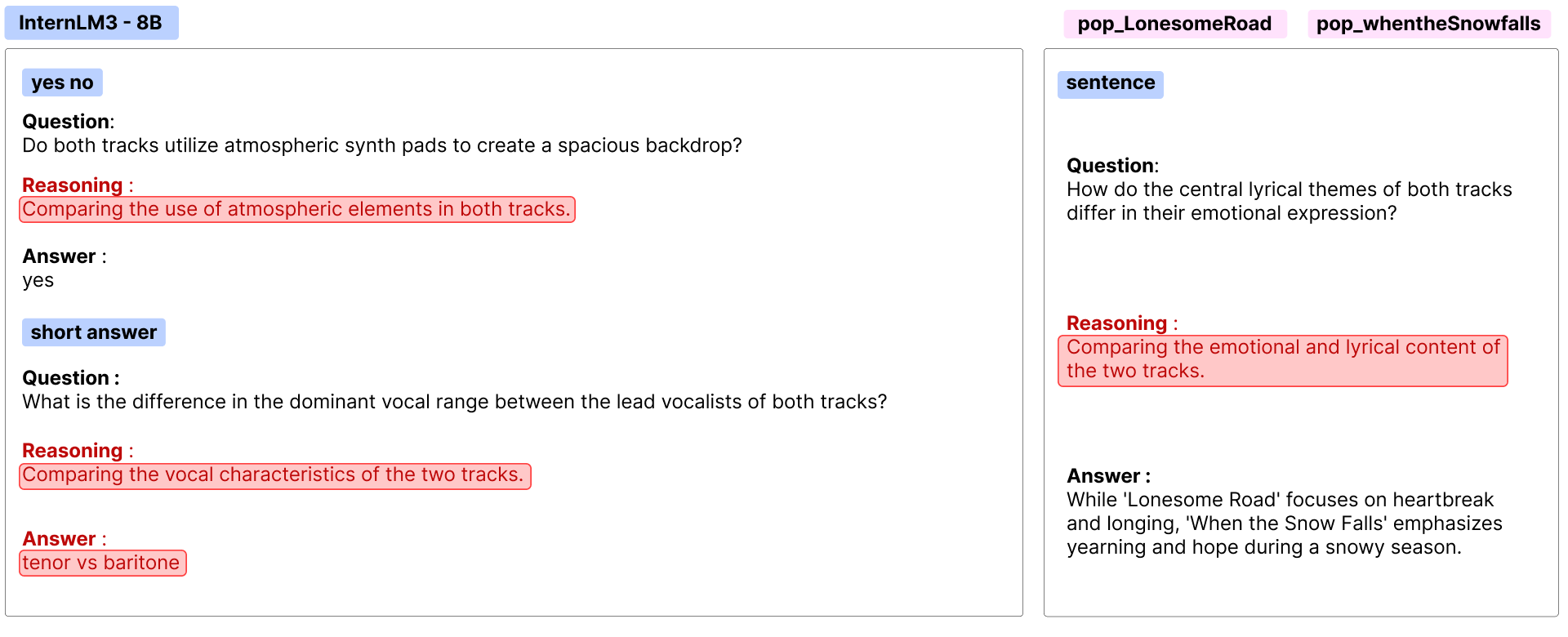}
\caption{
Qualitative example of Stage~3 comparative QA generation using \textbf{InternLM3-8B}.
The model frequently produces reasoning that closely resembles the question itself,
and we observe occasional incorrect answers in the short-answer type.
These issues indicate weaker grounding and reliability for comparative QA generation.
Red highlights denote incorrect or weakly supported outputs.
}
\label{fig:stage3_intern}
\end{figure*}

\begin{figure*}[t]
\centering
\includegraphics[width=\textwidth]{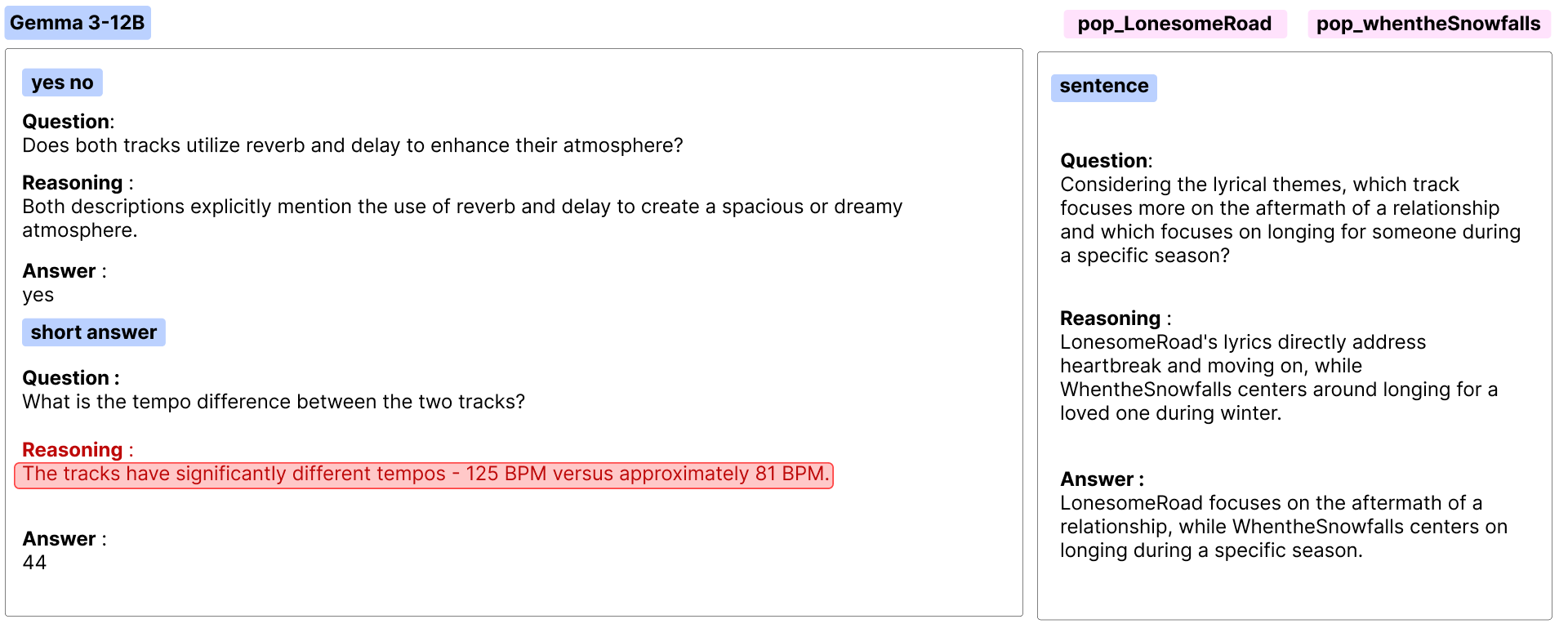}
\caption{
Qualitative example of Stage~3 comparative QA generation using \textbf{Gemma 3 12B}.
The model generally produces reasonable questions and answers; however, the
reasoning is sometimes less detailed or less step-wise, which reduces
interpretability for dataset construction and downstream supervision.
Red highlights indicate parts where the reasoning lacks sufficient specificity.
}
\label{fig:stage3_gemma}
\end{figure*}

\begin{figure*}[t]
\centering
\includegraphics[width=\textwidth]{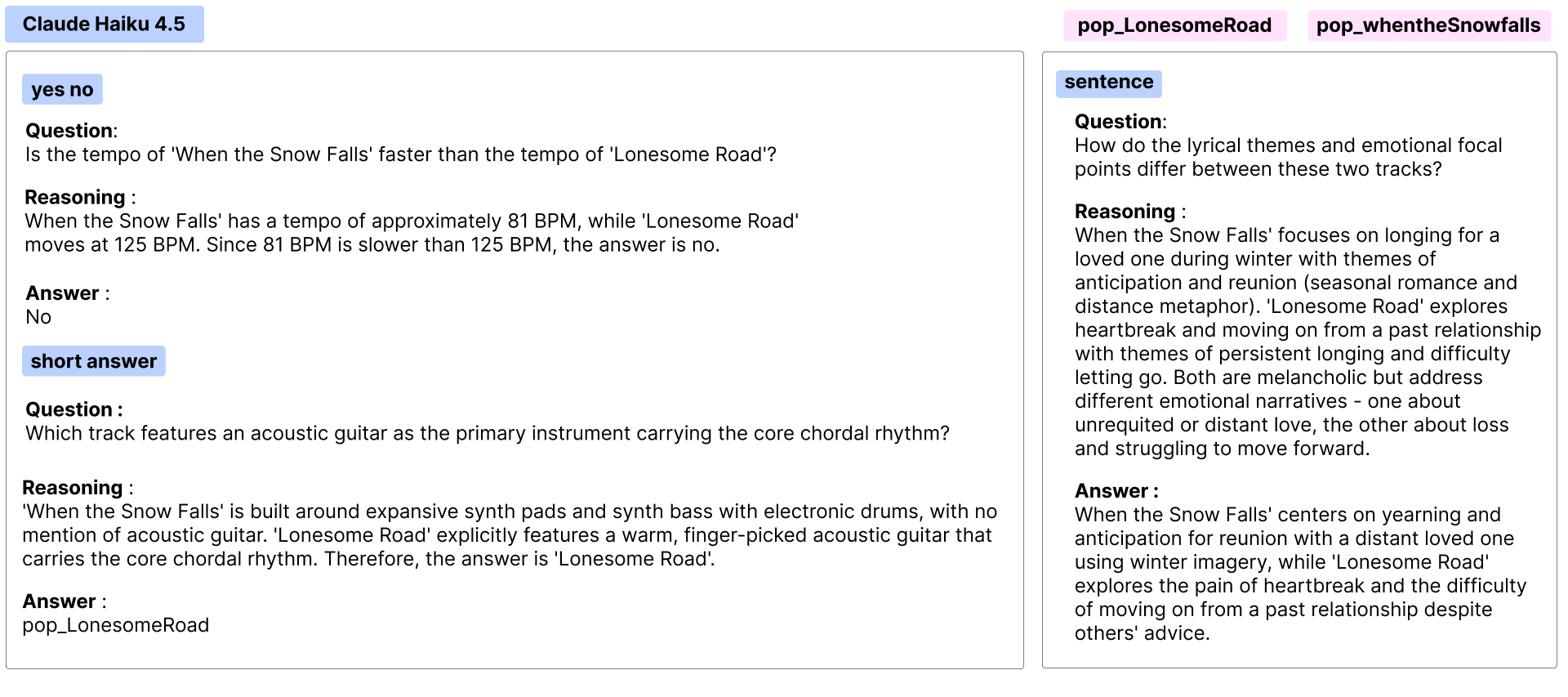}
\caption{
Qualitative example of Stage~3 comparative QA generation using \textbf{Claude 4.5 Haiku}.
The model typically produces strong comparative questions and fluent sentence-level
answers. However, compared to GPT-5 mini, the reasoning is less consistently
structured in explicit step-by-step form, which motivated our preference for
GPT-5 mini in final dataset construction.
}
\label{fig:stage3_haiku}
\end{figure*}

In Stage~3, we experimented with multiple LLM generators for comparative QA
construction before selecting \textbf{GPT-5 mini} as the primary generator.
To illustrate qualitative differences, Figures~\ref{fig:stage3_qwen},
\ref{fig:stage3_intern}, \ref{fig:stage3_gemma}, and \ref{fig:stage3_haiku}
show the three-question QA group (yes/no, short-answer, sentence-level)
generated for the \emph{same} audio pair across four alternative models.
Red highlights in these figures mark recurring issues that motivated our decision
to not use these models for final dataset construction.

\paragraph{Qwen3-32B.}
We observed that the short-answer question sometimes yields uncommon or subjective
word choices (e.g., stylistic adjectives rather than a strict selection), which can
break the intended answer format. In addition, sentence-level outputs often mirror
the reasoning text with minimal reformulation, reducing diversity between
\textit{reasoning} and \textit{answer}.

\paragraph{InternLM3-8B.}
The model frequently produces reasoning that closely repeats the question prompt,
and we occasionally observed incorrect selections in the short-answer type,
indicating weaker grounding and higher error rates for comparative decisions.

\paragraph{Gemma-3-12B.}
While overall quality is competitive, the reasoning is sometimes less detailed
or less step-wise than desired, which reduces interpretability and usefulness
for downstream training or analysis.

\paragraph{Claude-4.5-Haiku.}
This model generally produces strong comparative QA. However, \textbf{GPT-5 mini}
more consistently provides step-by-step, grounded reasoning, which we
found preferable for reliable dataset construction and for using the reasoning
field as a high-quality supervisory signal.

\section{Baseline Inference Settings}
\label{sec:appendix-baseline-settings}

\subsection{Metadata-Only Baseline}
\label{sec:appendix-metadata}

To verify that audio-derived information is genuinely needed, we tested three conditions on a stratified sample of 500 QA pairs using GPT-4o-mini (Table~\ref{tab:metadata_baseline}). Metadata-only accuracy (12.9\%) is worse than random guessing (34.3\%), while caption-only achieves 65.6\%---a +52.7pp gap ($p < 10^{-108}$). This confirms that the benchmark requires genuine music understanding beyond coarse metadata.

\begin{table}[ht]
\centering
\small
\begin{tabular}{lccc}
\toprule
\textbf{QA Type} & \textbf{Metadata} & \textbf{Caption} & \textbf{Random} \\
\midrule
Yes/No & 0.163 & 0.885 & 0.440 \\
Short Answer & 0.084 & 0.808 & 0.527 \\
Sentence (F1) & 0.141 & 0.276 & 0.062 \\
\midrule
Overall & 0.129 & 0.656 & 0.343 \\
\bottomrule
\end{tabular}
\caption{Metadata-only vs.\ caption-only vs.\ random baseline on 500 QA pairs. Metadata provides only genre, speed, and vocal gender.}
\label{tab:metadata_baseline}
\end{table}

\subsection{Generation Parameters}
Table~\ref{tab:gen_params} summarizes decoding parameters used for baseline inference.
Unless otherwise noted, we use the model's default tokenizer and stop conditions.

\begin{table}[t]
\centering
\small
\resizebox{\columnwidth}{!}{%
\begin{tabular}{lcccc}
\toprule
\textbf{Model} & \textbf{Temp.} & \textbf{Top-p} & \textbf{Top-k} & \textbf{Max Tokens} \\
\midrule
MU-LLaMA & 0.6 & 0.8 & -- & 512 \\
Music Flamingo & 0.0 (greedy) & -- & -- & 512 \\
Qwen3-Omni & 0.01 & 0.1 & 1 & 256 \\
GPT-4o-mini & 0.0 (greedy) & -- & -- & 512 \\
\bottomrule
\end{tabular}%
}
\caption{Decoding parameters used for baseline inference and LLM-as-a-Judge evaluation.}
\label{tab:gen_params}
\end{table}

\subsection{Prompt Templates}
\label{sec:appendix-prompts}

We evaluate two baseline setups (Figure~\ref{fig:baseline_setup}):
(i) \textit{caption-based} baselines, where each track is first converted into a textual description and an LLM answers comparative questions using the two captions; and
(ii) \textit{multi-audio} baselines, where models directly process two audio inputs along with the question.
To reduce prompt sensitivity, we use a single standardized template per setup, with only minimal model-specific formatting (e.g., special tokens for audio inputs).

\subsubsection{Caption-based comparison prompt.}
\begin{lstlisting}[style=promptstyle]
You are a music expert. Based on the following descriptions of two music tracks, answer the question.

=== Track A: {audio1_name} ===
{caption1}

=== Track B: {audio2_name} ===
{caption2}

Question: {question}

Answer:
\end{lstlisting}

\subsubsection{Multi-audio comparison prompt.}
\begin{lstlisting}[style=promptstyle]
[Audio 1] [Audio 2]
You are given two music tracks.
Track A: {audio1_name}
Track B: {audio2_name}

Question: {question}

Please compare the two tracks and answer the question concisely.
\end{lstlisting}

\paragraph{Caption prompts.}
For caption-based models, we prompt each model to produce a detailed single-track description.
For Music Flamingo, we request structured coverage (genre, tempo, key, instrumentation, vocals, production, mood).
For MU-LLaMA, we use a concise captioning instruction (``Describe this music in detail.'') and then apply the standardized caption-based comparison prompt above.

\begin{table*}[t]
\centering
\small
\setlength{\tabcolsep}{4pt}
\renewcommand{\arraystretch}{1.15}
\begin{tabularx}{\textwidth}{l l X X X}
\toprule
\textbf{Model} & \textbf{Error Type} & \textbf{Question (Sentence-level)} & \textbf{Reference} & \textbf{Model Output} \\
\midrule
GPT-4o-Audio & \textbf{(A)} Comp.Col. &
In one sentence, how do the vocal style and emotional tone differ between Track A and Track B? &
Track A features a restrained, airy vocal delivery with a more melancholic tone, while Track B has a fuller vocal presence and a warmer, more uplifting feel. &
Track A and Track B both present expressive vocals and emotional atmospheres. Track A leans into a softer, intimate delivery, while Track B maintains a similarly emotive style but with a slightly different texture overall. \\

GPT-4o-Audio & \textbf{(B)} Attr.Hal. &
In one sentence, compare the rhythmic groove and instrumentation between Track A and Track B. &
Track A has a steady mid-tempo groove with prominent electric guitars and live drums, whereas Track B uses a lighter rhythmic feel with more electronic percussion and synth layers. &
Track A features a hip-hop inspired beat with heavy bass and sampled percussion, while Track B relies on acoustic strumming and orchestral strings, creating a more cinematic rhythm section. \\
\midrule

Music Flamingo & \textbf{(B)} Attr.Hal. &
In one sentence, how do the primary instruments differ between Track A and Track B? &
Track A is driven by electric guitars and live drums, whereas Track B emphasizes synth textures and programmed percussion. &
Track A is led by piano and orchestral strings, while Track B centers on brass sections and saxophone riffs. \\

Music Flamingo & \textbf{(A)} Comp.Col. &
In one sentence, compare the mood and atmosphere of Track A and Track B. &
Track A feels darker and more introspective, while Track B is brighter and more upbeat. &
Both tracks create strong atmospheres and convey distinctive moods through their instrumentation and arrangement. \\
\midrule

MU-LLaMA & \textbf{(A)} Comp.Col. &
How do the key, tempo, instrumentation, and vocal delivery of each track create their distinct emotional atmospheres? &
Track A uses contrasting musical elements to produce a dramatic, cinematic feel, while Track B creates an uplifting trance atmosphere. &
Both Track A and Track B share an upbeat and energetic quality, but they create distinct emotional atmospheres through their instrumentation and vocal delivery. \\

MU-LLaMA & \textbf{(B)} Attr.Hal. &
In one sentence, how do the two tracks differ in genre, tempo, vocal timbre, and production approach? &
Track A is an introspective, guitar-driven Indie/Alternative Rock song at about 100 BPM with a melancholic male tenor. &
Track A has an indie genre with a weird and tribal feel, likely slower in tempo, while Track B is an upbeat electronic fusion of traditional Indian and Western styles. \\
\midrule

Qwen3-Omni & \textbf{(A)} Comp.Col. &
In one sentence, how do the two tracks compare in terms of overall mood, instrumentation, and lyrical themes? &
Track A is an upbeat, polished indie-folk pop track with lyrics about adventure, while Track B is a reflective acoustic piece about personal loss. &
While both tracks feature a prominent ukulele and have an uplifting, positive mood, they differ significantly in theme and instrumentation. \\

Qwen3-Omni & \textbf{(B)} Attr.Hal. &
Compare how tempo, harmonic complexity, production style, and instrumentation differ between the two tracks. &
Track A sits around 120 BPM with shifting feels and distorted guitars, while Track B is a slower, melodic rock track at 85 BPM. &
Track A is a slow, atmospheric piece with ambient textures, while Track B is a fast-paced electronic track with aggressive synths. \\
\bottomrule
\end{tabularx}
\caption{
Representative error examples for sentence-level comparative questions.
Each row shows one incorrectly answered item (LLM-judge score $<3$),
together with its assigned error type.
Comp.Col.: Comparative Collapse; Attr.Hal.: Attribute Hallucination.
}
\label{tab:error_examples}
\end{table*}

\end{document}